\documentclass[12pt]{article}
\usepackage{graphicx}

\begin{document}

\begin{center}
\Large {\bf  Massless Cosmic Strings in Spacetimes with Global Parabolic Isometries}\footnote{A talk given at International Bolgolyubov Conference 
``Problems of Theoretical and Mathematical Physics'', dedicated to the 110th anniversary of N.N.Bogolyubov’s birth,
Dubna-Moscow, 9-13.09.2019}
\end{center}

\bigskip
\bigskip

\begin{center}
D.V. Fursaev
\end{center}

\bigskip
%\bigskip

\begin{center}
{\it Dubna State University \\
     Universitetskaya str. 19\\
     141 980, Dubna, Moscow Region, Russia\\

  and\\

  the Bogoliubov Laboratory of Theoretical Physics\\
  Joint Institute for Nuclear Research\\
  Dubna, Russia\\}
 \medskip
\end{center}

\bigskip
\bigskip

\begin{abstract}
A class of curved spacetimes with global parabolic isometries (GPI) is introduced. These isometries have fixed point sets on two-dimensional null surfaces
which can be interpreted as worldsheets of massless cosmic strings. Back reaction effects of the  strings in such spacetimes
can be described exactly, in terms of a nontrivial holonomy at the worldsheet.  We show that the GPI spacetimes are type $N$ geometries of the Petrov classification.
We describe a number of features of these  spacetimes, including properties of Killing horizons associated to GPI.
As an example, we consider a circular massless cosmic string in the de Sitter universe and present the metric in new coordinates centered at the
string worldsheet.
\end{abstract}

\noindent
PACS: 11.27.$+$d, 98.80.Cq

\newpage

\section{Introduction}\label{intr}

The parabolic transformations, also known as null rotations, are a subgroup 
of the Lorentz group. The transformations leave invariant the quadratic form $-t^2+x^2+y^2$ and
planes of constant $t-x$. If $t,x,y,z$ are coordinates in the Minkowsky spacetime $R^{1,3}$, the parabolic transformations
$(x')^\mu=M^\mu_{~\nu}\left(\omega\right)x^\nu$  are defined as:
\begin{equation}\label{1.1}
u'=u~~,~~
v'=v+2\omega y+\omega^2u~~,~~
y'=y+\omega u~~,~~z'=z~~,
\end{equation}  
where $u=t-x$, $v=t+x$ are light-cone coordinates and $\omega$ is some real parameter. One can check that (\ref{1.1}) make a 
one-parameter group, $M(\omega_1)M(\omega_2)=M(\omega_1+\omega_2)$. We put $c=1$ for the velocity of light.

Transformations (\ref{1.1})  do not change the metric 
\begin{equation}\label{1.2}
ds^2=-dudv+dy^2+dz^2~~.
\end{equation} 
so we call (\ref{1.1}) the global parabolic isometry (GPI) of $R^{1,3}$. The important feature of GPI is that 2-dimensional null surface $\cal S$,
$u=y=0$, is the fixed point set of the isometry.  The surface $\cal S$ can be interpreted as the worldsheet of
a massless cosmic string (MCS),  a one-dimensional object of zero thickness which is stretched along 
the $z$ axis and moves along $x$ axis with the speed of light. MCS in a flat spacetime can be obtained from 
common massive cosmic strings \cite{Kibble:1976sj} as a limiting case,  when the velocity of the string reaches the speed of light, mass tends to zero, while energy  remains finite \cite{Barrabes:2002hn}. As a result of this limit, a holonomy along a closed contour around a massive string 
is transformed into a non-trivial holonomy around the MCS \cite{vandeMeent:2012gb}. The holonomy of MCS is an element of the parabolic subgroup 
described above. Therefore, similarly to massive strings MCS allow global gravitational effects. The effects look as mutual transformations
of trajectories of massive bodies or light rays, when the string moves in between two trajectories.

A method how to describe physical effects around massless cosmic strings in a flat spacetime
by using GPI  has been developed in \cite{Fursaev:2017aap}.  Massless strings generate perturbations of the velocities of bodies resulting  in overdensities of matter. 
The strings also shift energies of photons, and may yield additional 
anisotropy of cosmic microwave background, if we consider MCS in a cosmological context. These effects of MCS are direct
analogs  of, respectively,  wake effects \cite{Brandenberger:2013tr} and the Kaiser-Stebbins effect \cite{Stebbins:1987va}, \cite{Sazhina:2008xs} 
known for common cosmic strings.  

An example of a curved spacetime with GPI is a de Sitter spacetime.  MCS in the de Sitter universe are circles of the horizon radius.
The physical effects of such strings have been described in \cite{Fursaev:2018spa}.  The aim of the present paper is to go further. We
introduce a general class of curved spacetimes with global parabolic isometries and outline their basic properties.
This class of geometries allows one an exact description of the backreaction and, hence, gives a tool to study physical effects caused by MCS 
in terms of the parabolic holonomy.

The paper is organized as follows.   GPI spacetimes are defined in Sec. \ref{s1}. The isometries demand that
these spacetimes are type $N$ geometries of the Petrov classification.  In Sec. \ref{Killing} we show how to construct a locally GPI spacetime
with a nontrivial holonomy at a null 2D hypersurface $\cal S$. We use an analog of GPI coordinates  (\ref{1.1}).  In this Section we also discuss properties of Killing horizons  corresponding to parabolic isometries. By using  the GPI coordinates we discover a new metric for a de Sitter spacetime with a circular MCS,
see Sec. \ref{newsolution}. The results of the work are briefly discussed in Sec. \ref{sum}.

\section{Definitions and basic properties}\label{s1}
\setcounter{equation}0

Transformations (\ref{1.1}) were introduced as isometries of the Minkowsky spacetime $R^{1,3}$. Now we consider an arbitrary 
curved manifold $\cal M$ and a coordinate chart with coordinates $u,v,y,z$ on $\cal M$. As in case of flat spacetime 
(\ref{1.2}),  we assume that $u=C$, $v=C'$, with arbitrary real constants $C$ and $C'$, 
are intersecting null hypersurfaces which make a foliation of $\cal M$ (the so called  double 
null foliation, see \cite{Brady:1995na}). The intersections of the null hypersurfaces are codimension 2 spacelike 2-surfaces $\cal B$.
We denote coordinates on $\cal B$ by $y$ and $z$.  The metric on $\cal M$ can be written as \cite{Brady:1995na}
\begin{equation}\label{1.3}
ds^2=-2e^\lambda dudv+h_{ab}(dx^a+s^a du+\bar{s}^a dv)(dx^b+s^b du+\bar{s}^b dv)~~,
\end{equation} 
where $x^1=y,x^2=z$ and $\lambda, s^a, \bar{s}^a$ are some functions of coordinates. One of the `vectors', $s^a$ or $\bar{s}^a$, can be eliminated
by a coordinate transformation.

We say that $\cal M$ has a  global parabolic isometry if metric (\ref{1.3}) is invariant with respect to (\ref{1.1}). We also demand that components
of (\ref{1.3}) are regular in the limit when the fixed point set of the isometry, $u=y=0$, is approached.  The requirement of GPI restricts
(\ref{1.3}) to the following form:
\begin{equation}\label{1.4}
ds^2=-2e^\lambda (dudv-dy^2)+h(dz+s du)^2~~.
\end{equation} 
Here $\lambda$, $h$ and $s$ are functions of $u$, $z$ and the invariant
\begin{equation}\label{1.5}
\theta=uv-y^2~~.
\end{equation} 

Let $\zeta^\mu$ be the Killing vector field which generates GPI.  In the given coordinates 
it has the components $\zeta^u=0, ~\zeta^v=2y,~\zeta^y=u,~\zeta^z=0$.
One can introduce the tetrads:
\begin{equation}\label{1.6}
n^\mu=\sqrt{2}e^{-\lambda/2}\delta^\mu_v~~,~~l^\mu=\sqrt{2}e^{-\lambda/2}(\delta^\mu_v-s\delta^\mu_z)~~,~~
q^\mu=\sqrt{2}e^{-\lambda/2}\delta^\mu_y~~,~~p^\mu={1 \over \sqrt{h}}\delta^\mu_z~~,
\end{equation} 
which obey the normalization conditions:
\begin{equation}\label{1.7}
l^2=n^2=(l\cdot p)=(l\cdot q)=(n\cdot p)=(n\cdot q)=(p\cdot q)=0~~,~~(l\cdot n)=-2~~,~~p^2=q^2=1~~.
\end{equation} 
The null vectors $n$ and $l$ are normal vectors to the hypersurfaces $u=C$, $v=C'$, respectively.  Under infinitesimal coordinate transformations
$\delta x^\mu=\omega \zeta^\mu$ the tetrads change as $\delta e^a=-{\cal L}_\zeta e^a$. It can be checked that
\begin{equation}\label{1.8}
\delta n=0~~,~~\delta l=2\omega q~~,~~\delta p=0~~,~~\delta q=\omega n~~.
\end{equation}
(\ref{1.8}) are equivalent to finite transformations $(e')^a=M^a_{~b}(\omega)e^b$, 
\begin{equation}\label{1.9}
n'=n~~,~~l'=l+2\omega q+\omega^2 n~~,~~p'=p~~,~~q'=q+\omega n~~.
\end{equation}
That is $M^a_{~b}(\omega)$ are parabolic Lorentz transformations introduced in (\ref{1.1}).

The Lie derivatives along $\zeta$ of tensor structures on $\cal M$ must vanish.  By using the Newman-Penrose formalism one can write
the Weyl tensor of $\cal M$  in the following form:
\begin{equation}\label{1.10}
C_{\mu\nu\lambda\rho}=\Psi_4 ~V_{\mu\nu}V_{\lambda\rho}+c.c.~~,
\end{equation}
where $c.c.$ denotes complex conjugation, $\Psi_4=\Psi_4(\theta,u,z)$ is one of the Weyl scalars, see \cite{Chandrasekhar:1985kt}, and
\begin{equation}\label{1.11}
 V_{\mu\nu}=n_\mu m_\nu-m_\mu n_\nu~~,~~m_\mu=p_\mu+i q_\mu~~.
\end{equation}
The invariance of $C_{\mu\nu\lambda\rho}$ results from the transformation law $m'=m+i\omega n$. Relation (\ref{1.10}) implies that
other four complex Weyl scalars on $\cal M$ vanish. Therefore, GPI spacetimes belong to the $N$ type of the Petrov 
classification \cite{Chandrasekhar:1985kt}.

By using similar arguments one can fix the structure of a stress-energy tensor on $\cal M$
\begin{equation}\label{1.12}
 T_{\mu\nu}=a_1~g_{\mu\nu}+a_2~n_\mu n_\nu+a_3~p_\mu p_\nu+a_4~(n_\mu p_\nu+p_\mu n_\nu)~~,
\end{equation}
where $a_k$ are functions of invariants  $\theta,u,z$. Thus, $\cal M$ can be Ricci-flat (all $a_k=0$) and it can be a spacetime with a cosmological constant
($a_k=0$, except $a_1=C$). In a more general case, (\ref{1.12}) obeys the condition $T_{\mu\nu}n^\mu n^\nu=0$.

\section{The string, the backreaction, and the horizon}\label{Killing}
\setcounter{equation}0

We identify the worldsheet $\cal S$ of a massless cosmic string with a fixed point set of the Killing vector field $\zeta=2y\partial_v+u\partial_y$, which 
generates GPI. It is clear that $\cal S$ is defined by equations $u=y=0$, and it is a null two-dimensional
surface.  Vectors $n$ and $m$ taken at $\cal S$ are two normal vectors of the worldsheet. $n$ is also a tangent vector to $\cal S$, together with $p$.
One can interpret $n$ at $\cal S$ as the 4-velocity of the string. One can check with the help of (\ref{1.6}) that $\nabla_n n=0$, that is each point of the string moves along a lightlike geodesics.

The parabolic isometries leave invariant all null hypersurfaces $u=C$, however the hypersurface $u=0$ is special since
it includes $\cal S$. We denote the hypersurface $u=0$ by $\cal H$  and call it {\it the string  horizon}.
As was shown in \cite{Fursaev:2017aap}, \cite{Fursaev:2018spa}, the string horizon $\cal H$
plays an important role in constructing from $\cal M$ a spacetime $\tilde{\cal M}$ which takes into account backreaction 
effects caused by the MCS with worldsheet $\cal S$. Due to the isometry the local geometry of $\tilde{\cal M}$ is identical to that
of  $\cal M$  except a set of singular points on $\cal S$, where $\tilde{\cal M}$ has a non-trivial holonomy. The holonomy
is determined by an element of parabolic isometry where the parameter $\omega$, see (\ref{1.1}), is related to the energy of the string.  
This situation
is somewhat similar to the case of manifolds with conical singularities where holonomies are related to spatial rotations and depend on the 
string tension.

The null surface $\cal S$ divides $\cal H$ into two parts, ${\cal H}_+$ ($y>0$) and  ${\cal H}_-$  ($y<0$).  According to \cite{Fursaev:2017aap}, \cite{Fursaev:2018spa},  $\tilde{\cal M}$ can be constructed  if one cuts $\cal M$ along, say,  ${\cal H}_-$, and applies to the tangent space 
the required parabolic transformation when crossing the cut.  This procedure yields a self-consistent method to describe, for example, mutual 
rotation of two trajectories which pass the string from different sides.

Intersection of $\cal H$ and null hypersurfaces $v=C'$  are codimension 2 spacelike 2-surfaces $\cal B$, see Sec. \ref{s1}.
One can say that $\cal H$ is foliated by a family of $\cal B$. Parabolic transformations (\ref{1.1}) change $v$ and the foliation. It is important
that in the considered problem all sections $\cal B$ for any type of the foliation have the same metrics: 
\begin{equation}\label{2.1}
dl^2=h_{yy} dy^2+h_{zz}dz^2~~,
\end{equation}
where $h_{yy}=2e^\lambda$, $h_{zz}=h$, and, as follows from (\ref{1.4}), (\ref{1.5}), $h_{ab}=h_{ab}(y^2,z)$. The string position
at $\cal B$ is $y=0$. It is easy to see
that the line $y=0$ is a geodesic on $\cal B$.
At $\cal B$ one can define a `length of the string' as an integral  $L=\int \sqrt{h} dz$. The length of the
string does not depend on coordinate $v$, it is constant on $\cal S$. The string preserves its form during the motion.

The string horizon has a number of important properties related to the global isometries.  

1. It follows from the definition that $\zeta^2=u^2\partial^2_y=u^2e^\lambda/2$,  $\zeta$ is everywhere spacelike except $\cal H$. The Killing vector field is null, $\zeta^2=0$, on $\cal H$.  Therefore,  $\cal H$ is the Killing event horizon.

2.  On the horizon, the Killing vector is directed along the normal vector, $\zeta\sim n$, see (\ref{1.9}), if $y\neq 0$.

3.  $\cal S$ is a fixed point set of the Killing field, $\zeta=0$ on $\cal S$.

4. One can check that $\nabla_\zeta\zeta=0$ on $\cal H$.  This means, that  $\cal H$ is an extremal Killing horizon with respect to 
the given isometry.

Under certain conditions $\cal H$ may coincide with a common Killing horizon. Let us assume that  $\lambda$, $h$ and $s$  
in (\ref{1.5}) are functions of $\theta$ and $z$ (i.e., they do not depend explicitly on $u$).  Then components of  (\ref{1.5}) are invariant under the transformations:
\begin{equation}\label{2.2}
 u'=e^{\beta}u~~,~~ v'=e^{-\beta}v~~,
\end{equation}
where $\beta$ is a constant. Transformations (\ref{2.2}) are boosts generated by the Killing vector field $\eta=u\partial_u-v\partial_v$. One can easily check 
that:

1. $\eta^2=2uv e^\lambda$, thus, $\eta$ can be timelike, spacelike  or null. $\eta^2=0$ and $\eta \sim n$ on $\cal H$. Hence, $\cal H$ is the Killing horizon with respect to $\eta$. 

2. Also $\nabla_\eta \eta=-\eta$ on $\cal H$, that is $\cal H$ has a nonzero surface gravity with respect to $\eta$.

3. Finally, simple algebra shows that $[\eta,\zeta]=\zeta$. 

Note that boosts (\ref{2.2}) leave invariant form of (\ref{1.1}) but 
change the parameter $\omega$ to $e^{-\beta}\omega$.  The corresponding parameter in the holonomy at the string worldsheet changes in the same way.
Physical interpretation of this fact is that energy of the string, like the energy of a photon, depends on the motion of an observer.

An example of GPI spacetime with two Killing fields will be presented in next Section.

\section{Metric of the de Sitter spacetime with MCS}\label{newsolution}
\setcounter{equation}0

The global parabolic isometry of the de Sitter geometry was used \cite{Fursaev:2018spa} to describe physical effects caused
by a circular MCS moving along the cosmological horizon. Here we analyze this example along the lines of Sec. \ref{s1} and Sec. \ref{Killing}.
We discover a new form of the de Sitter metric.

The de Sitter can be embedded in a five-dimensional Minkowsky spacetime $R^{1,4}$ (with coordinates $X^K$),
\begin{equation}\label{3.1}
-X_0^2+X_1^2+X_2^2+X_3^2+X_4^2=-UV+Y^2+X_2^2+X_3^2=1~~.
\end{equation}
Here $U=X_0-X_1$,  $V=X_0+X_1$, $Y=X_4$. For simplicity we put the de Sitter radius equal to unity. Consider in $R^{1,4}$ a null hypersurface $U=Y=0$, 
which can be interpreted as worldsheet of a massless brane. Its intersection 
with (\ref{3.1}) is a null surface in (\ref{3.1}). It is a space product $R^1\times S^1$ whose cross sections are circles $X_2^2+X_3^2=1$. Points on the circles move along null geodesics on the de Sitter spacetime. 

The coordinate transformations $X'=M(\omega)X$ from the parabolic subgroup of $O(1,4)$, which leave (\ref{3.1}) and the 
surface $U=Y=0$ invariant, are
\begin{equation}\label{3.2}
U'=U~~,~~
V'=V+2\omega Y+\omega^2 U~~,~~
Y'=Y+\omega U~~,~~X_2'=X_2~~,~~X_3'=X_3~~.
\end{equation} 

Define the following coordinates $u,v,y,z$:
\begin{equation}\label{3.3}
U={2u \over 1-\theta}~~,~~V={2v \over 1-\theta}~~,~~Y={2y \over 1-\theta}~~,~~X_2={1+\theta \over 1-\theta}\cos z~~,~~
X_3={1+\theta \over 1-\theta}\sin z~~,
\end{equation} 
where $\theta$ is defined by (\ref{1.5}).  We assume that $\theta \neq 1$. The de Sitter metric takes the new simple form
\begin{equation}\label{3.4}
ds^2={1 \over (1-\theta)^{2}}\left(-4(dudv-dy^2) +(1+\theta)^{2}dz^2\right)~~,
\end{equation} 
with points $z$ and $z+2\pi$ identified.

It is easy to see that transformations (\ref{3.2}) induce parabolic transformations (\ref{1.1}) and GPI is explicit in 
(\ref{3.4}) (compare with (\ref{1.3})).

The null hypersurface $U=0$ intersects the de Sitter spacetime at $u=0$ (which is the string horizon $\cal H$), the hypersurface $Y=0$ intersects at $y=0$.
The string worldsheet is at $u=y=0$. One can check by using (\ref{3.4}) that sections $\cal B$ of the horizon are 2-spheres.

Metric (\ref{3.4}) demonstrates that the cosmological horizon $\cal H$ in the de Sitter universe is the Killing horizon for two Killing fields, see Sec. \ref{Killing},
$\eta=u\partial_u-v\partial_v$ and $\zeta=2y\partial_v+u\partial_y$. The field $\eta$ is well known, it corresponds to boosts and is characterized by non-vanishing  surface gravity. The field $\zeta$ generates parabolic transformations and has vanishing surface gravity on $\cal H$.

\section{Discussion}\label{sum}

The aim of this work was to introduce a new type of spacetimes which possess global parabolic isometries.
Such spacetimes are interesting since fixed point sets of the isometries can be considered as 
worldsheets of massless cosmic strings, and backreaction effects caused by strings can be described in terms of parabolic holonomies  along the lines of our 
approach developed in \cite{Fursaev:2017aap},\cite{Fursaev:2018spa} .

One of the surprises of our analysis is the discovery of a new simple metric element of the de Sitter universe.
The metric is given in terms of GPI coordinates with the `origin' at the string worldsheet.
If massless cosmic strings appear in the early universe they can be observed in the same way as their massive cousins.  It is 
important that experimental data, such as CMB anisotropy, allow one to discriminate MCS form massive cosmic strings \cite{Fursaev:2018spa}.
New metric (\ref{3.4})  may be a useful tool to study physical effects caused by MCS in the de Sitter universe, see Sec. \ref{intr}.
Studying properties of (\ref{3.4}) is an interesting research subject.

Another research direction is finding other examples of GPI spacetimes introduced in Sec. \ref{s1}. 
Concerning mathematical aspects of Killing horizons, as we saw in this work they may have hidden symmetries. There may be an additional
independent Killing field which is null on the horizon.
Common Killing horizons in quantum theory are known to result in important effects related to thermodynamics.
It would interesting to analyze how these properties change when motion of the string along the horizon creates
singularities.

\bigskip
\bigskip
\bigskip

%\newpage

\end{document}